\documentclass[prb,aps,twocolumn,floatfix,longbibliography]{revtex4-2}
\usepackage[utf8]{inputenc}
\usepackage{natbib}
\usepackage{graphicx}
\usepackage{xcolor}
\usepackage[tbtags]{amsmath}
\usepackage{amssymb}
\usepackage{gensymb}
\usepackage[normalem]{ulem}
\usepackage{array}

\newcommand{\sgn}{{\rm sgn\,}}

\renewcommand{\vec}[1]{{\bf #1}}
\renewcommand{\hat}[1]{{\bf {\widehat #1}}}
\renewcommand{\phi}{\varphi}
\renewcommand{\epsilon}{\varepsilon}

\renewcommand{\dag}{\dagger}

\begin{document}

\title{{Visualizing structure of correlated ground states using collective charge modes}}

\author{Micha{\l} Papaj}
\thanks{These authors contributed equally to this work.}
\affiliation{Department of Physics, University of California, Berkeley, CA 94720, USA}

\author{Guangxin Ni}
\address{National High Magnetic Field Laboratory, Tallahassee, Florida, 32310, USA}
\address{Department of Physics, Florida State University, Tallahassee, Florida 32306, USA}

\author{Cyprian Lewandowski}
\thanks{These authors contributed equally to this work.}
\address{National High Magnetic Field Laboratory, Tallahassee, Florida, 32310, USA}
\address{Department of Physics, Florida State University, Tallahassee, Florida 32306, USA}

\begin{abstract}
The variety of correlated phenomena in moir\'e systems is incredibly rich, spanning effects such as superconductivity, a generalized form of ferromagnetism, or even charge fractionalization. This wide range of quantum phenomena is partly enabled by the large number of internal degrees of freedom in these systems, such as the valley and spin degrees of freedom, which interplay decides the precise nature of the ground state.  Identifying the microscopic nature of the correlated states in the moir\'e systems is, however, challenging, as it relies on interpreting transport behavior or scanning-tunneling microscopy measurements. Here we show how the real-space structure of collective charge oscillations of the correlated orders can directly encode information about the structure of the correlated state, focusing in particular on the problem of generalized Wigner crystals in moiré transition metal dichalcogenides. Our analysis builds upon our earlier result [10.1126/sciadv.adg3262] that the presence of a generalized Wigner crystal modifies the plasmon spectrum of the system, giving rise to new collective modes. We focus on scanning near-field optical microscopy technique (SNOM), fundamentally a charge-sensing-based method, and introduce a regime under which SNOM can operate as a probe of the spin degree of freedom.
\end{abstract}

\maketitle

\section{Introduction}

Moir\'e materials have emerged as the ideal platform to explore the correlated physics \cite{Balents2020, 10.1038/s41563-020-00840-0, makSemiconductorMoireMaterials2022}. The variety of correlated phenomena in moir\'e systems is incredibly rich, including effects such as superconductivity \cite{cao2, luSuperconductorsOrbitalMagnets2019b, chenSignaturesTunableSuperconductivity2019, Yankowitz1059}, a generalized form of ferromagnetism\cite{Nuckolls2023,2023arXiv230410586K}, or even charge fractionalization\cite{Park2023,Cai2023,Zeng2023}. This wide range of quantum phenomena is partly enabled by the large number of internal degrees of freedom in these systems, such as the valley and the spin. The two-dimensional (2D) nature of the materials combined with the gate tunability allows for experimental studies of these systems using various modern experimental probes beyond the transport paradigm\cite{10.1038/s41567-021-01438-2,doi:10.1126/science.abc2836,Ideue2021, Jiang2019,Kerelsky2019,Choi2019,PhysRevLett.129.117602,PhysRevLett.129.147001,2022arXiv220914322X,2022arXiv220912964Z,2022arXiv221101352Z}. This fact is significant in the moir\'e systems as they exhibit high spatial inhomogeneity\cite{Grover2022}, necessitating exploration using local probe techniques. However, such probes are typically only sensitive to the local density of states, thus not allowing for a straightforward imaging of the internal degrees of freedom.

\begin{figure*}[t]
\includegraphics[width=\linewidth]{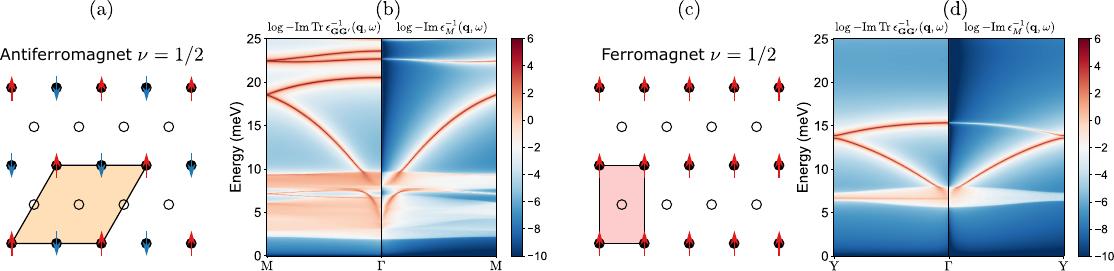}

\caption{Structure and plasmon spectra of generalized Wigner crystals at filling fraction $\nu=1/2$. (a) AFM Wigner crystal with magnetic moments out of plane. The Wigner crystal unit cell is 4 times larger than moir\'e unit cell. (b) Comparison between the trace of inverse dielectric function matrix and macroscopic dielectric function. Some of the modes are absent in the macroscopic function. (c) FM Wigner crystal with magnetic moments out of plane. The Wigner crystal unit cell is 2 times larger than moir\'e unit cell and changes from hexagonal to rectangular. (d) Comparison between the trace of inverse dielectric function matrix and macroscopic dielectric function. Both modes are visible in the macroscopic function.}
\label{fig:fig_1}
\end{figure*}

In this work, we build on our results of Ref. \cite{Papaj2023}, where we show how the presence of a generalized Wigner crystal\cite{tangSimulationHubbardModel2020, reganMottGeneralizedWigner2020, liImagingTwodimensionalGeneralized2021, xuCorrelatedInsulatingStates2020} modifies the plasmon spectrum of the system, giving rise to new collective modes. The number of the new modes directly corresponds to the enlargement of the unit cell, which provides information about the spin structure of the correlated ground state provided that the spin structure of a candidate ground state is known, c.f. Fig. \ref{fig:fig_1}. As such, observing a specific number of folded plasmon modes would, in principle, allow one to ascertain the viability of different candidate ground states~\cite{panQuantumPhaseDiagram2020, wuHubbardModelPhysics2018,PhysRevB.104.075150,2022arXiv220900664L} directly. In the current manuscript, we further show that the spin degree of freedom can be accessed via the real-space structure of the plasmon modes. Therefore, we propose using scanning near-field optical microscopy techniques\cite{10.1038/nature11254, 10.1038/nature11253, PhysRevLett.119.247402, hesp2019collective}, fundamentally a charge-sensing-based method, to operate as a probe of the spin degree of freedom.

The field of plasmonics in 2D systems \cite{10.1038/nphys2615, 10.1038/s41586-018-0136-9, basovPolaritonsVanWaals2016,In2022}, graphene in particular, usually assumes the approximation of modeling \cite{PhysRevB.75.205418,Wunsch_2006,PhysRevB.84.045429,PhysRevB.80.245435} the system as a homogeneous electronic liquid as typical wavelengths of the surface-plasmon polariton modes exceed interatomic spacing, ${\sim\,}$0.1 nm. In moir\'e systems, however, this paradigm changes as the effective unit cell of the material is ${\sim\,}$10 nm~\cite{PhysRevLett.99.256802,Bistritzer12233,PhysRevB.95.075420}, thus opening avenues for exploration of consequences of interaction strength enhancement \cite{PhysRevLett.125.066801} and local field effects \cite{adlerQuantumTheoryDielectric1962, wiserDielectricConstantLocal1963}  - that is, the consideration of electric field modulation on the length of the moiré unit cell. The topic of local field effects in conventional solids is well-studied in the literature\cite{wiserDielectricConstantLocal1963,adlerQuantumTheoryDielectric1962,PhysRevB.73.045112,PhysRevB.35.5585,Baldereschi1979}, particularly in the context of optical absorption and high-precision ab initio simulations. As we will discuss, the implicit assumption in most theoretical works on local field effects restricts them to the large wavelength approximation. In this work, in addition to serving as a concrete experimental blueprint for visualization of the spin structure of generalized Wigner crystals, we expand on the theoretical framework for treating dielectric response with local field effects, focusing on finite momentum properties of the dielectric response and highlighting how it connects to modern experimental probes that allow for finite-momentum study of 2D systems.

Our manuscript is structured as follows: Section IIa-IIc summarizes the properties of the dielectric function with local field effects included, the toy model of a generalized Wigner crystal for the WSe$_2$/WS$_2$ moiré heterostructures as well as the main results of Ref. \cite{Papaj2023}. Section IId-f introduce the framework necessary to describe the two common experimental platforms, far-field optical absorption and scanning near-field optical microscopy, and analyze the conditions under which collective modes can be excited using these experimental methods. Section IIf concludes by considering specific applications of these techniques to the generalized Wigner crystal candidate ground states in  WSe$_2$/WS$_2$ moiré heterostructures and demonstrate how a real-space spin structure can be readily visualized using scanning near-field optical microscopy.

\section{Results \& Methods}
\subsection{Dielectric function with local field effects}

We begin by reviewing the formalism for treating local field effects introduced in the seminal works of Ref.~\cite {wiserDielectricConstantLocal1963,adlerQuantumTheoryDielectric1962}. We follow the notation and reproduce parts of the discussion presented in our earlier work of Ref.\cite{Papaj2023} for consistency. We assume that a crystal has a periodic structure that gives rise to a set of reciprocal lattice vectors $\vec{G}$. Consequently, the real space dielectric function $\epsilon(\vec{r},\vec{r'})$\cite{wiserDielectricConstantLocal1963,adlerQuantumTheoryDielectric1962,PhysRevB.35.5585} can be expressed via a Fourier transform
\begin{equation}
\epsilon(\vec{r},\vec{r'}) = \frac{1}{\Omega} \sum_{\vec{q},\vec{G},\vec{G'}} e^{i(\vec{q}+\vec{G})\cdot \vec{r}} \epsilon_{\vec{G}\vec{G'}}(\vec{q}) e^{-i(\vec{q}+\vec{G'})\cdot \vec{r'}}\,,
\end{equation}
where $\Omega$ denotes the unit cell area. If the crystal is translationally invariant, $\epsilon(\vec{r},\vec{r'})=\epsilon(\vec{r}-\vec{r'})$ and we can carry out summation over one set of reciprocal lattice vectors $\vec{G'}$. This procedure yields an effective Fourier transform over a single continuous momentum $\vec{q}_{con}$ that we split into an integral over momentum $\vec{q}$ inside the 1st Brillouin zone (BZ) and summation over reciprocal momenta $\vec{G}$, such that $\vec{q}_{con}=\vec{q}+\vec{G}$. We thus see that for $\vec{G}=0$ and the contribution $\vec{q}$ inside the 1st BZ gives rise to modulation of the dielectric function on lengthscales larger than the unit cell. Correspondingly, the inclusion of higher Fourier harmonics $\vec{G}\neq 0$ gives rise to real space modulation of the dielectric function on the length scales smaller than the unit cell. Despite the matrix structure of the Fourier-transformed dielectric function, the relation between the fields due to free and all charges holds\cite{adlerQuantumTheoryDielectric1962,wiserDielectricConstantLocal1963}. That is, the displacement field due to the free charges is related to the total electric field by $\vec{D}_{\vec{G}}(\mathbf{q}, \omega) = \sum_{\vec{G'}} \epsilon_{\vec{G}\vec{G'}}(\vec{q}, \omega) \vec{E}_{\vec{G'}}(\mathbf{q}, \omega)$, where $\vec{E}_{\vec{G'}}(\mathbf{q}, \omega)$ are the Fourier components of the total electric field in the crystal. The individual $\vec{G}\neq \vec{G'} \neq 0$ components of the dielectric function matrix do not have a straightforward interpretation\cite{adlerQuantumTheoryDielectric1962,wiserDielectricConstantLocal1963,PhysRevB.35.5585}, beyond the argument that they describe screening on lengthscales of the unit cell. We stress that the above Fourier decomposition is a statement about Fourier transforms and periodicity of the underlying crystal, thus remains valid no matter the approximation schemes used to compute the system's dielectric function, which we discuss below.

The Refs.~\citenum{wiserDielectricConstantLocal1963,adlerQuantumTheoryDielectric1962} shows that when calculated within the random phase approximation (RPA) and accounting for the periodicity of the crystal, the dielectric function matrix takes the following form \cite{wiserDielectricConstantLocal1963,adlerQuantumTheoryDielectric1962, PhysRevB.73.045112}:
\begin{align}
    \epsilon_{\mathbf{G}\mathbf{G'}}(\mathbf{q}, \omega) =&\delta_{\mathbf{G}\mathbf{G'}}-T_{\vec{G}\vec{G'}}(\vec{q}, \omega)\label{eq:epsGG_def} \\
      T_{\vec{G}\vec{G'}}(\vec{q}, \omega)=&  V_{\mathbf{q} + \mathbf{G}} \sum_{n,m, \mathbf{k}} \frac{f_0(\epsilon_{n\mathbf{k}}) - f_0(\epsilon_{m\mathbf{k}+\mathbf{q}})}{\omega + i 0^+ + \epsilon_{n\mathbf{k}} - \epsilon_{m\mathbf{k} + \mathbf{q}}}
     \times   \nonumber\\
     &\times \eta_{\vec{q},\vec{G}}^{nm}(\vec{k})^* \eta_{\vec{q},\vec{G'}}^{nm}(\vec{k}) \label{eq:pol_GG_def}
\end{align}
Here $ V_\mathbf{q} = \frac{2 \pi e^2}{\kappa|\mathbf{q}|}$ is the Fourier transform of the Coulomb potential, $\kappa$ is the dielectric constant of the surrounding insulating gate material, $f_0(\epsilon) = (e^{\beta(\epsilon-\mu)}+1)^{-1}$ is the Fermi-Dirac distribution, with $\mu$ and $\beta = 1/k_B T$ the chemical potential and the inverse temperature, respectively. The state overlap $\eta_{\vec{q},\vec{G'}}^{nm}(\vec{k})$
\begin{align}
\label{eq:wavefunction_def}
   \eta_{\vec{q},\vec{G}}^{nm}(\vec{k}) = \frac{1}{\Omega} \int_{\Omega} d^2 \vec{r} ~u_{n \vec{k}}(\vec{r})^\dag e^{-i \vec{G}\cdot \vec{r}} u_{m \vec{k}+\vec{q}} (\vec{r})
\end{align}
is determined by $u_{n \vec{k}}(\vec{r})$, the cell-periodic part of the Bloch wave function $\psi_{n \vec{k}}(\vec{r}) = u_{n \vec{k}}(\vec{r}) e^{i \vec{k} \cdot \vec{r}}$ defined for an eigenstate from band $n$ with a crystal momentum $\vec{k}$ and energy $\epsilon_{n \vec{k}}$. The integration in Eq.~\eqref{eq:wavefunction_def} is over the unit cell with area $\Omega$ in real space. The usual form of the RPA dielectric function can be recovered by taking only the $\vec{G}=\vec{G'}=0$ component, i.e., neglecting the variations of the dielectric function on the distances smaller than the unit cell size.

\subsection{Plasmon dispersion}

A plasmon is a collective oscillation of charge density\cite{PhysRev.33.195,PhysRev.85.338,PhysRev.92.609,mahan2000many-particle,2008qtelbookG}, enabled through the Coulomb interaction of electrons. To determine the dispersion of plasmons, one seeks sustained electric-field oscillation in the absence of free charges ($\vec{D}_{\vec{G}}(\mathbf{q}, \omega) = 0$). In the case of the dielectric matrix introduced before (See also Supplemental Materials), it is given by the solution of a zero eigenvalue problem for $\epsilon_{\vec{G}\vec{G'}}(\mathbf{q}, \omega)$ with the characteristic polynomial given by $\text{det}\, \epsilon_{\vec{G}\vec{G'}}(\mathbf{q}, \omega) = 0$. If the local field effects are neglected, and one considers only long wavelength oscillations, the plasmon dispersion reduces to the conventional form\cite{PhysRevB.75.205418,Wunsch_2006,PhysRevB.84.045429,PhysRevB.80.245435} $\epsilon_{00}(\mathbf{q}, \omega) = 0$ where the matrix element $\epsilon_{00}(\mathbf{q}, \omega)$ has a dependence on $\omega$ and BZ momentum $\vec{q}$. An example of the plasmon modes determined by the solution of this determinant equation are shown in Fig. \ref{fig:fig_1}, for the microscopic model introduced in the following section. 

A practically relevant question, and one of the motivations for this manuscript, is whether all of the plasmon modes given by $\text{det}\, \epsilon_{\vec{G}\vec{G'}}(\mathbf{q}, \omega) = 0$ can be experimentally excited. In the later part of the manuscript, we focus in particular on one experimental technique, scanning near-field optical microscopy\cite{10.1038/nature11254, 10.1038/nature11253, PhysRevLett.119.247402, hesp2019collective}, but here we mention the guiding physical interpretation of the zero eigenvalue eigenvectors of the characteristic plasmon equation above. The relevant eigenvectors correspond to the real-space pattern of charge oscillations, which produce an oscillating transverse magnetic wave on the surface of the 2D materials (See discussion in Supplemental Materials). As we will see, however, the experimental accessibility of some of these new plasmon branches may depend on the specific measurement technique used.

\subsection{Generalized Wigner Crystal Model}
We pause briefly to describe our platform of choice for the analysis - the WSe$_2$/WS$_2$ heterobilayers with zero twist angle \cite{makSemiconductorMoireMaterials2022, reganMottGeneralizedWigner2020, tangSimulationHubbardModel2020, liImagingTwodimensionalGeneralized2021, xuCorrelatedInsulatingStates2020}. Here again, we reproduce for consistency the discussion of our earlier work Ref. \cite{Papaj2023}, inviting an interested reader to consult it for more details and extended motivation of the modelling. To describe the low energy theory of the WSe$_2$/WS$_2$ heterobilayers, we use the Bistrizer-MacDonald-type continuum model \cite{Bistritzer12233}.  The valence moir\'e bands of this platform are comprised of WSe$_2$ hole pockets centered around K and K' points of the Brillouin zone. Due to the spin-valley locking of TMDs \cite{xiaoCoupledSpinValley2012}, the K/K' valley degree of freedom can be identified with the electron spin, and therefore the system is two-fold degenerate. The charge neutrality point of the entire structure lies within the bulk gap of WSe$_2$, and WS$_2$, hence the top-most doubly degenerate valence moir\'e band {(with the narrow bandwidth of below 10 meV)} corresponds to the filling $-2 \le \nu \le 0$ as defined in the experiments \cite{xuCorrelatedInsulatingStates2020}. To study the fillings $0 \le \nu \le 2$, one can also use a comparable model, which after appropriate adjustments to the effective mass describes moir\'e conduction bands formed by WS$_2$ electron pockets.

Within each valley, we approximate the Hamiltonian $H_0$ by a parabolic band with periodic moir\'e potential $V_M(\mathbf{r})$ that results from the insulating layer of WS$_2$ placed on top of it \cite{wuHubbardModelPhysics2018, zhangMoirQuantumChemistry2020}
 \begin{equation}
 \label{eq:ham_kinetic}
     H_0 = - \frac{\hbar^2 k^2}{2 m^*} + V_M(\mathbf{r}),
 \end{equation}
 with $m^* = 0.472\,m_e$ being the effective mass of the hole pocket of WSe$_2$. The moir\'e potential can be decomposed in terms of its Fourier components within the first shell of reciprocal lattice vectors $\mathbf{b}_i$, with $\mathbf{b}_1 = 2\pi/a_M[2/\sqrt{3}, 0]$, and the remaining five vectors obtained by $\pi/3$ rotations. Here $a_M = a/\delta \approx 8.2$ nm is the moir\'e lattice constant with $a = 0.328$ nm and $\delta=4\%$ the lattice constant of WSe$_2$ and lattice mismatch between WSe$_2$/WS$_2$, respectively \cite{wuHubbardModelPhysics2018, zhangMoirQuantumChemistry2020}. The moir\'e potential can therefore be expanded as
\begin{equation}
    V_M(\mathbf{r}) = \sum_j V_{\mathbf{b}_j} e^{i \mathbf{b}_j \cdot \mathbf{r}},
\end{equation}
where $V_{\mathbf{b}_1} = V_0 e^{i \psi}$, with $V_0 = 15\,\mathrm{meV}$ and $\psi=45\degree$ for WSe$_2$/WS$_2$ \cite{zhangMoirQuantumChemistry2020}, and the remaining coefficients are obtained from symmetry since $V_\mathbf{b} = V_{R(2\pi/3, \mathbf{b})}$ and $V_\mathbf{b} = V^*_{-\mathbf{b}}$ (here $R(\theta,\mathbf{b})$ denotes counterclockwise rotation of vector $\mathbf{b}$ by angle $\theta$). The moir\'e potential breaks down the parabolic band given by Eq.~\eqref{eq:ham_kinetic} into a set of mini-bands defined within the moiré Brillouin zone.

When a generalized Wigner crystal state forms, it is pinned to a fraction of the moir\'e lattice sites, with the periodicity of the Wigner crystal lattice dependent on the exact form of the correlated ground state \cite{panBandTopologyHubbard2020, panQuantumPhaseDiagram2020,PhysRevB.104.075150}. This translates to varying sets of reciprocal lattice vectors $\mathbf{G}$, shorter than the moir\'e reciprocal lattice (i.e., moir\'e BZ becomes folded). Microscopically, the emergence of the correlated order and the enlargement of the unit cell is captured within the framework of a triangular lattice Hubbard model \cite{panQuantumPhaseDiagram2020, wuHubbardModelPhysics2018,PhysRevB.104.075150,2022arXiv220900664L}. Depending on the strength of interactions and filling of the bands, the ground state can, for example, be ferromagnetic or antiferromagnetic with various patterns of in-plane or out-of-plane magnetic moments, c.f. Fig. \ref{fig:fig_1}. Building on the observation that the Hartree-Fock ground states have a generalized Wigner crystal character\cite{wuHubbardModelPhysics2018,panQuantumPhaseDiagram2020} with charge mainly localized around selected moir\'e potential minima, we use an effective potential that enforces a spin-structure consistent with the orientation of the magnetic moments as shown in the self-consistent calculations. Specifically, in terms of the vectors $\mathbf{G}$, we take an effective potential that enforces the spin texture predicted by Hartree-Fock calculations \cite{panBandTopologyHubbard2020, panQuantumPhaseDiagram2020} as given by:
\begin{equation}
\label{eq:spin_potential}
    H_M^{\mathbf{G}\mathbf{G}'}(\mathbf{k}) = V_B \sum_\mathbf{\tau} e^{-i (\mathbf{G}-\mathbf{G}')\cdot \mathbf{\tau}}  \mathbf{B}(\mathbf{G}-\mathbf{G}', \mathbf{\tau})\cdot \mathbf{\sigma}
\end{equation}
Here $V_B$ is a fitting parameter, $\mathbf{B}(\mathbf{G}-\mathbf{G}', \mathbf{\tau})$ is a vector that determines the orientation of the effective magnetic field at a given basis site $\mathbf{\tau}$ of the crystal, and $\mathbf{\sigma}$ is a vector of Pauli matrices for the spin degree of freedom. This approach, combined with the enlargement of the unit cell due to the Wigner crystal formation and using experimental estimates for the correlated gaps magnitude \cite{xuCorrelatedInsulatingStates2020}, allows us to give an overview of the qualitative features of the dynamic dielectric response of the system that can enable us to identify the structure of the correlated ground state.

\subsection{Plasmon folding \& challenges in experimental detection}

The central result of Ref. \cite{Papaj2023} relies on the idea that the correlated order enlarges the effective unit cell of the moiré system. As a result, the longitudinal charge collective mode of the original metallic system, the plasmon, develops a gap in its spectrum and becomes folded an integer number of times. The number of folded branches precisely corresponds to the enlargement factor of the unit cell of the Wigner crystal, therefore acting as a qualitative signature of the candidate ground state\cite{panQuantumPhaseDiagram2020, wuHubbardModelPhysics2018,PhysRevB.104.075150,2022arXiv220900664L}, c.f. Fig. \ref{fig:fig_1}a-d. Effectively, through the dependence of the enlarged unit cell size on the spin structure, the number of the plasmon branches thus serves as a crude measurement of the spin order.

The above prediction, although conceptually relatively straightforward to check, poses several experimental challenges. Specifically, the challenges are related to the requirement of a wide frequency range of the experimental probes (all in the low THz range) and a requirement for a probe to create a plasmon with finite momentum. The tunability of low-frequency sources is limited; thus, typically, a measurement in a low THz range operates at few selected frequencies - as such, continuously sweeping a frequency range to count the plasmon resonances, as depicted in Fig.~\ref{fig:fig_1} is not straightforward. The complexity of such an optical experiment would further be exacerbated by the requirement that the optical probe effectively imparts a finite momentum to the plasmon. In the following few paragraphs, we focus on these two complications to address how the proposed folded plasmon modes can be observed.

To circumvent the limited number of finite momentum and low-frequency experimental probes, we want to first study whether far-field observation methods that irradiate a material measuring its optical conductivity could excite the folded plasmon modes. In typical 2D materials, longitudinal charge oscillations are gapless; thus, they disappear at momentum $\vec{q}\to 0$. The new folded collective modes shown in Fig. ~\ref{fig:fig_1}b,d (left panels) demonstrate, however, that the folded plasmons remain gapped as $\vec{q}\to 0$. As such, it is instructive to understand whether far-field optical probes can detect these modes.

To assist with the analysis, we introduce a toy model that helps to understand the properties of the folded plasmon modes. Specifically, we consider the dielectric matrix $\epsilon_{\vec{G}\vec{G'}}(\vec{q},\omega)$ truncating its dimensions to include only the two nearest reciprocal lattice vectors along some high symmetry axis in each direction, i.e. $\vec{G},\vec{G'}=-2\vec{G_1},-\vec{G_1},0,\vec{G_1},2\vec{G_1}$ making the dielectric matrix $5\times 5$. Unlike the toy model introduced in Ref.\cite{Papaj2023}, which was $2\times2$, here we require the included vectors to span a larger range to capture the structure of the eigenvectors, as we will see in what follows. The truncated dielectric matrix $\tilde{\epsilon}_{\vec{G}\vec{G'}}(\vec{q},\omega)$ takes the form
\begin{equation}
\label{eq:tilde_mat}
    \tilde{\epsilon}_{\vec{G}\vec{G'}}(\vec{q},\omega) \equiv \begin{pmatrix}
\epsilon_{-2-2} & \epsilon_{-2-1} & \epsilon_{-2 0} & \epsilon_{-2 1} & \epsilon_{-2 2} \\
\epsilon_{-1-2} & \epsilon_{-1-1} & \epsilon_{-1 0} & \epsilon_{-1 1} & \epsilon_{-1 2} \\
\epsilon_{0-2} & \epsilon_{0-1} & \epsilon_{0 0} & \epsilon_{0 1} & \epsilon_{0 2} \\
\epsilon_{1-2} & \epsilon_{1-1} & \epsilon_{1 0} & \epsilon_{1 1} & \epsilon_{1 2} \\
\epsilon_{2-2} & \epsilon_{2-1} & \epsilon_{2 0} & \epsilon_{2 1} & \epsilon_{2 2}
\end{pmatrix} .
\end{equation}
In the above expression for conciseness we suppress the arguments $\omega$, $\vec{q}$ in each matrix entry. We also label the row and column entries with integers $m,n=-2,-1,0,1,2$ corresponding to the integer multiples of the reciprocal lattice vectors as argued earlier.
The diagonal terms correspond to the dielectric function computed in neighboring BZ cells in an extended zone scheme. In the absence of the local field effects, $\tilde{\epsilon}_{\vec{G}\neq\vec{G'}}(\vec{q},\omega) \equiv 0$, the condition
$\det \tilde{\epsilon}_{\vec{G}\vec{G'}}(\vec{q},\omega)=0$ translates to $\tilde{\epsilon}_{\vec{G}\vec{G}}(\vec{q},\omega)=0$. Correspondingly $\tilde{\epsilon}_{00}(\vec{q},\omega_{0})=0$ would give the lowest plasmon dispersion $\omega_{0}(\vec{q})$ (i.e. the conventional plasmon in a metal). In contrast, the equations $\tilde{\epsilon}_{\vec{G}\vec{G}=\vec{G_1}}(\vec{q},\omega_{+1})=0$ and $\tilde{\epsilon}_{\vec{G}\vec{G}=-\vec{G_1}}(\vec{q},\omega_{-1})=0$ would correspond to the first folded branches $\omega_{\pm 1}(\vec{q})$. The above discussion applies irrespective of the microscopic approximation scheme used to determine the form of the dielectric matrix elements.

To understand the impact and behavior of the off-diagonal elements of $\tilde{\epsilon}_{\vec{G}\vec{G'}}(\vec{q},\omega)$ it is helpful to focus on a particular approximation scheme for computing the dielectric function. We use the RPA approximation given in Eq.~\eqref{eq:epsGG_def}. The off-diagonal terms, due to the dependence on the shifted Bloch wavefunctions, are typically suppressed exponentially with the difference $\vec{G}-\vec{G'}$ as a consequence of the continuum model structure (e.g. see Refs. \cite{Ishizuka2021,2023arXiv230707531K} and the discussion therein). This serves as a useful theoretical tool allowing us to keep track of the order of the off-diagonal terms, which we do by introducing a parameter $\alpha$ to each entry $m,n$ in the matrix $\tilde{\epsilon}_{\vec{G}\vec{G'}}(\vec{q},\omega)$ in Eq.~\eqref{eq:tilde_mat} $\epsilon_{mn} \to \alpha^{|m-n|}\epsilon_{mn}$. Using this perturbation technique, it is possible to compute the determinant and inverse of  $\tilde{\epsilon}_{\vec{G}\vec{G'}}(\vec{q},\omega)$ order by order in powers of $\alpha$ (see Supplemental materials), which we will use in the analysis that follows.

\subsection{Far-field optical conductivity measurement}

We now return to the analysis of the experimental signatures. An optical conductivity measurement is carried out by irradiating a sample and measuring either the transmitted or reflected intensity of light (e.g. see \cite{PhysRevLett.101.196405} for a typical experiment). Measured quantities are related to the materials' optical conductivity, denoted as $\sigma(\omega)$. To arrive at the relation between $\sigma$ and dielectric matrix $\epsilon_{\vec{G}\vec{G'}}(\vec{q},\omega)$ with local field effects, we consider the continuity equation \cite{2008qtelbookG}
\begin{equation}
\frac{\partial \rho}{\partial t} + \nabla \cdot \vec{j} = 0\,,
\end{equation}
where $\rho$ denotes time and position dependent charge density and $\vec{j}$ denotes the time and position dependent current density. The continuity equation allows to relate the physical (causal) density-density response function $\Pi_{\vec{G}\vec{G'}}(\vec{q},\omega)$ to the physical current-current response function $\sigma_{\vec{G}\vec{G'}}(\vec{q},\omega)$:
\begin{equation}
 \sigma_{\vec{G}\vec{G'}}(\vec{q},\omega) = -i \frac{ e^2 \omega}{(\vec{q}+\vec{G})\cdot (\vec{q}+\vec{G'})} \Pi_{\vec{G}\vec{G'}}(\vec{q},\omega)
\end{equation}
Here we stress that $\Pi_{\vec{G}\vec{G'}}(\vec{q},\omega)$ is the physical (causal) density-density response function which can be exactly expressed in terms of the irreducible (proper) $\tilde{\Pi}_{\vec{G}\vec{G'}}(\vec{q},\omega)$ density-density response function
\begin{align}
&\Pi_{\vec{G}\vec{G'}}(\vec{q},\omega) = -\sum_{\vec{G''}} \tilde{\Pi}_{\vec{G}\vec{G''}}(\vec{q},\omega) (\epsilon^{\mathrm{exact}}(\vec{q},\omega))^{-1}_{\vec{G''}\vec{G'}}\\
&\epsilon^{\mathrm{exact}}_{\vec{G}\vec{G'}}(\vec{q},\omega) = 1-V_{\vec{q}+\vec{G}} \tilde{\Pi}_{\vec{G}\vec{G''}}(\vec{q},\omega)\,,
\end{align}
where $\tilde{\Pi}_{\vec{G}\vec{G'}}(\vec{q},\omega)$ describes the response to the screened potential and is defined diagrammatically as the sum of all the ``proper'' diagrams \cite{2008qtelbookG}. We introduced also an expression for the exact dielectric matrix function $\epsilon^{\mathrm{exact}}_{\vec{G}\vec{G'}}(\vec{q},\omega)$. Approximating $\tilde{\Pi}_{\vec{G}\vec{G'}}(\vec{q},\omega)$ as the bare propagator bubble, c.f. Eq.\eqref{eq:pol_GG_def},
\begin{equation}
\tilde{\Pi}_{\vec{G}\vec{G'}}(\vec{q},\omega)=\tilde{\Pi}^0_{\vec{G}\vec{G'}}(\vec{q},\omega) \equiv T_{\vec{G}\vec{G'}}(\vec{q},\omega)/V_{\vec{q}+\vec{G}}
\end{equation} 
gives the RPA approximation. Optical conductivity corresponds to the $\vec{q}_{con}\to 0$ limit, i.e. $\vec{q}\to 0, \vec{G}=\vec{G'}=0$,
\begin{equation}
\label{eq:optical_conductivity}
 \sigma_{00}(\vec{q}\to 0,\omega) = i \frac{ e^2 \omega}{q^2} \sum_{\vec{G}} \tilde{\Pi}^0_{0\vec{G}}(\vec{q},\omega) \epsilon^{-1}_{\vec{G} 0}(\vec{q},\omega)\,,
\end{equation}
where $\epsilon^{-1}_{\vec{G} \vec{G'}}(\vec{q},\omega)$ denotes the $\vec{G} \vec{G'}$ entry of the inverse of the RPA dielectric matrix of Eq. \eqref{eq:epsGG_def}.

It is instructive to consider the structure of the optical conductivity, Eq.\eqref{eq:optical_conductivity} in more detail. First, we notice that the expression involves an inverse of the dielectric matrix; thus, through the matrix inversion process, local field terms (off-diagonal entries in $\epsilon_{\vec{G}\vec{G'}}(\vec{q},\omega)$) modify the $0,0$ component of the inverted matrix. In the absence of local field effects, matrix inversion is trivial and reduces to the conventional 2D electron gas expression for optical conductivity. Lastly, we comment that the structure of the optical conductivity is related to another object frequently employed in describing local field effects - the macroscopic dielectric function $\epsilon_M(\vec{q},\omega)$\cite{wiserDielectricConstantLocal1963,adlerQuantumTheoryDielectric1962,PhysRevB.73.045112} defined as
\begin{equation}
\frac{1}{\epsilon_M(\vec{q},\omega)} \equiv \epsilon^{-1}_{00}(\vec{q},\omega)\,,
\end{equation}
where again $\epsilon^{-1}_{00}(\vec{q},\omega)$ denotes the $00$ entry of the inverse of the RPA dielectric matrix of Eq. \eqref{eq:epsGG_def}.
The expression for the macroscopic dielectric function is derived by asking what scalar (i.e., neglecting the matrix structure) dielectric function best approximates the dielectric response of the system in the limit of $\vec{q}\to 0$ (see Supplemental Material for derivation and Ref.\cite{wiserDielectricConstantLocal1963,adlerQuantumTheoryDielectric1962}). Light absorption of a material is then estimated by evaluating the loss function obtained using the macroscopic dielectric function as we do in Fig.\ref{fig:fig_1}b,d (right panels). Although loss function computed from the macroscopic dielectric function does not exactly correspond to the optical conductivity, due to the dependence on the inverse of the dielectric matrix, it captures the role local field effects play in the same fashion (see Supplemental Materials for further discussion).

\begin{figure*}[!t]
\includegraphics[width=\linewidth]{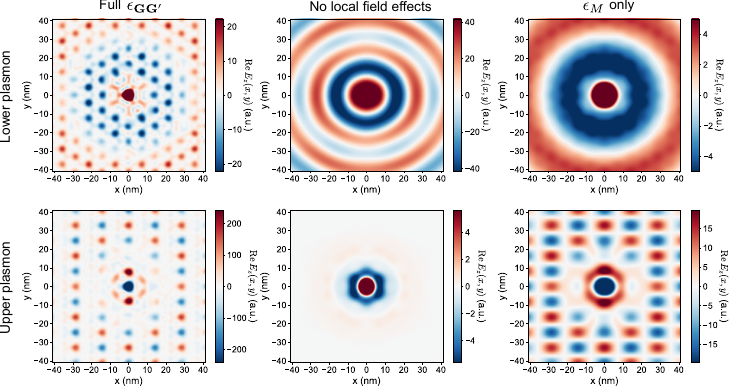}
\caption{Comparison between real part of electric field $z$ components as obtained through our full dielectric matrix approach (left column), approximation neglecting local field effects (middle column), and using macroscopic dielectric function only (right column), for two plasmon branches of AFM $\nu=1/2$ state. For the lower plasmon neglecting local field effects results in incorrect plasmon wavelength and using only $\epsilon_M$ misses the modulation on the moire\'e lengthscale. For the upper plasmon, without local field effects there is no excitation at all, and using $\epsilon_M$ only changes the moir\'e length scale modulation pattern. Here $z_0 = 0.5 L_M = 4.1$ nm and $L_M=8.2$ nm.}
\label{fig:fig_2}
\end{figure*}

We can formally evaluate matrix inverse in the expression of Eq.\eqref{eq:optical_conductivity} to arrive at
\begin{equation}
 \sigma_{00}(\vec{q}\to 0,\omega) = i \frac{ e^2 \omega}{q^2 \det\epsilon} \sum_{\vec{G}} \tilde{\Pi}^0_{0\vec{G}}(\vec{q},\omega) C^{T}_{\vec{G} 0}(\vec{q},\omega)\,,
\end{equation}
where $C^{T}_{\vec{G} \vec{G'}}(\vec{q},\omega)$ and $\det\epsilon$ are the matrix of cofactors and the determinant of the dielectric matrix (Eq. \eqref{eq:epsGG_def}) for a given $\omega,\vec{q}$, respectively.  This expression explicitly shows that the optical conductivity includes information about all of the collective plasmon modes due to the dependence on the determinant $\det\epsilon$, which vanishes at plasmon resonances. In practice, however, the matrix of co-factors includes zeros of the dielectric matrix as well, which, for $\vec{q}\to0 $ limit, as we will show, cancel the plasmon resonances in optical conductivity. This result is manifest in the toy $5\times5$ model, where we find that the optical conductivity schematically given by (we suppress $\vec{q}$ and $\omega$ for clarity) :
\begin{align}
&\frac{-i q^2}{e^2 \omega} \sigma_{00} = \frac{\tilde{\Pi}^0_{00}}{\epsilon _{00}}+\frac{\alpha^2}{\epsilon _{-1-1} \epsilon _{00}^2 \epsilon _{11}}\times  \\
&\times \left(-\tilde{\Pi}^0_{01} \epsilon _{-1-1} \epsilon _{00} \epsilon _{10}+\tilde{\Pi}^0_{00} \epsilon _{-1-1} \epsilon _{01} \epsilon _{10}+\right. \nonumber\\
&+\left.\tilde{\Pi}^0_{00} \epsilon _{-10} \epsilon _{0-1} \epsilon _{11}-\tilde{\Pi}^0_{0-1} \epsilon _{-10} \epsilon _{00} \epsilon _{11}\right)+O\left(\alpha^4\right) \nonumber
\end{align}
The expression yields a resonance in optical response whenever the zeros of the determinant of the dielectric matrix in the denominator are not eliminated by the zeros of the matrix of cofactors in the numerator. By analyzing the structure of the optical conductivity above, we find that plasmon resonance, say for the $\epsilon_{-1-1}$ entry, vanishes if the $\epsilon _{-10}$ dielectric matrix vanishes. This is always true since the off-diagonal local-field terms given by the RPA polarization function indeed must disappear for $\vec{q}\to 0$ due to the orthogonality of the eigenstates and, unlike in the terms $\vec{G}=0$, since the Fourier transform of the Coulomb potential $V_{\vec{G}}$ is finite there is no divergence. Therefore plasmon resonances, despite being gapped at $\vec{q}\to 0$ do not manifest in the optical conductivity. This can also be seen by analyzing the matrix inverse operation for a matrix in which entire column apart from the diagonal element is zero. In such a case, the corresponding diagonal element for the same column in the inverted matrix is just the inverse of the original diagonal element. Thus the local field effects that are represented by the off-diagonal elements of dielectric matrix do not influence the matrix inverse, and in turn the optical response, in the $\mathbf{q}\to 0$ limit.

Vanishing of the plasmon resonances in the optical conductivity also manifests in the structure of the eigenvectors of the dielectric matrix corresponding to the folded plasmon resonances. As argued previously, these eigenvectors, $\vec{E}_{\vec{G'}}(\mathbf{q}, \omega)$, represent the electric field normal modes corresponding to the self-sustained oscillations, e.g. $\sum_{\vec{G'}} \epsilon_{\vec{G}\vec{G'}}(\vec{q}, \omega) \vec{E}_{\vec{G'}}(\mathbf{q}, \omega) = 0$. We find, using the toy model and reproducing it in the simulations, that for $\vec{q}\to 0$, the eigenvectors, schematically, take the following form (here we keep only an order of magnitude in $\alpha$ for brevity and provide full expression in Supplemental Materials)
\begin{align}
(\vec{E}^0_{\vec{G}})^T &\sim (\alpha^2, \alpha, 1, \alpha, \alpha^2)\,,\nonumber\\
(\vec{E}^{1}_{\vec{G}})^T &\sim (\alpha^3, \alpha^2, 0, 1, \alpha)\,,\\
(\vec{E}^{-1}_{\vec{G}})^T &\sim (\alpha, 1, 0, \alpha^2, \alpha^3)\,,\nonumber
\end{align}
where the indices $s = 0,\pm 1$ in $\vec{E}^s_{\vec{G}}$ correspond to the original acoustic branch and the first folded branches. The $\vec{G}=0$ component vanishes for the folded branches, corresponding to the $E_{\vec{G} = 0}$ component of the electric field. As the Fourier decomposition of monochromatic light in the $\vec{q}_\mathrm{cont}\to0$ limit has support only in the $E_{\vec{G} = 0}$ as a result, there is no coupling between light and the charge oscillations.

In summary, although the folded plasmons are gapped at zero momentum, far-field optical conductivity measurement experiments would not be able to couple to these modes. These results indicate the necessity for experiments that effectively can excite plasmons with a finite momentum. Several successful techniques exist, which involve patterning gratings in devices, where the grating size sets the finite momentum \cite{Rodrigo2017,nnano.2012.59,10.1038/nphoton.2013.57,Ju2011}; irradiation with far-field light metallic contacts that consequently launch plasmons with a finite momentum set by the contact size \cite{Woessner2014,10.1038/s41586-018-0136-9}, or scanning near-field optical microscopy measurements wherein the AFM tip is irradiated and, effectively, gives rise to a dipole source. We focus here on the latter two approaches due to the relative ease of implementation.

\subsection{Scanning near-field optical microscopy measurement}

Scanning near-field optical microscopy\cite{10.1038/nature11254, 10.1038/nature11253, PhysRevLett.119.247402, hesp2019collective} is a complex measurement technique involving AFM tip and laser-irradiation, allowing for nanostructure investigation through breaking the far-field resolution limit by exploiting the properties of evanescent waves. From this paper's point of view, however, its key aspect is that it allows launching and detection of plasmons with a finite momentum. In a ``launch and detect'' mode of operations, the AFM tip is irradiated with an electromagnetic wave of a fixed frequency, effectively making the AFM tip act as an electric field dipole. Correspondingly, plasmons are launched radially (mirroring the symmetry of the AFM), propagating in all directions. When such a plasmon encounters an interface or a domain wall, part of its electromagnetic wave is reflected to the AFM, which can be detected, and by moving the AFM tip, effectively the electric field intensity that comprises the plasmon can be imaged\cite{10.1038/nature11254, 10.1038/nature11253}. Another mode of SNOM measurement experiment involves launching plasmons using a contact, which is later detected using the SNOM tip\cite{Woessner2014,10.1038/s41586-018-0136-9}. Purely geometrically, as we will see in what follows, the first mode corresponds to launching a circularly symmetric wave, while the second mode corresponds to a plane wave solution.

\begin{figure*}[t]
\includegraphics[width=\linewidth]{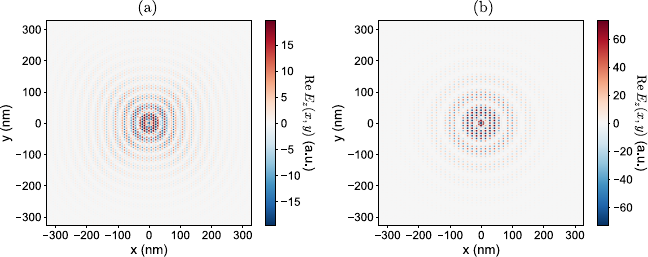}
\caption{Modulation of the plasmonic wave for longer distances demonstrated through $\mathrm{Re} E_z$ for (a) ferromagnet and (b) antiferromagnet. For the folded, upper plasmon branches the oscillations on the moir\'e length scale have opposite phase for each subsequent ring, which radius is determined by the Brillouin zone wave vector $\mathbf{q}$.}
\label{fig:fig_3}
\end{figure*}

A SNOM can be succinctly described as a time-dependent dipole electric potential\cite{PhysRevLett.125.256804}. Specifically, we imagine placing the dipole at a distance $z_0$ over the TMD surface. The oscillating dipole potential induces a time-dependent screening electric field inside the material - provided that the frequency of oscillations matches that of the plasmons, the SNOM tip can excite the collective modes. Explicitly, we model the dipole potential (See. Ref. \citenum{PhysRevLett.125.256804} and discussion in Supplemental Materials)
\begin{equation}
\phi_{dip}(\vec{r}_{3D}) = \frac{e^{i\omega t}}{\epsilon_r} \frac{\vec{P}\cdot(\vec{r}_{3D}-\vec{r}_0)}{(\vec{r}_{3D}-\vec{r}_0)^3}\,,
\end{equation}
where $\vec{P}=P \hat{z}$ is the dipole moment oriented along the $z$-axis perpendicular to the sample, $\vec{r}_{3D}=(\vec{r},z)$ is the position vector in 3D, and $\vec{r}_0 = (0,z_0)$ is the position of the SNOM tip above the sample. This potential produces a total potential in the sample (in momentum space, we suppress time dependence):
\begin{align}
\label{eq:phi_tot}
\phi_{tot}(\vec{q}+\vec{G}) &= \sum_{\vec{G'}} \epsilon^{-1}_{\vec{G}\vec{G'}}(\vec{q}) \phi_{dip}(\vec{q}+\vec{G})\\
\phi_{dip}(\vec{q}) &= -\frac{4 \pi P e^{i\omega t} }{\epsilon_r} e^{-q z_0}
\end{align}
A SNOM does not detect electric potential, but rather the electric field intensity. Considering the $z-$direction of the field, we find the final expression for the total electric field in the material to be given by (See Supplemental Materials for more details of the derivation)
\begin{align}
\label{eq:etot_snom}
E_{tot}(\vec{r},z) &= \sum_{\vec{q},\vec{G}} (-i(\vec{q+G})+\sgn(z) |\vec{q}+\vec{G}|~\hat{z})\times\\
&\times~\phi_{tot}(\vec{q}+\vec{G}) e^{i \vec{q}\cdot\vec{r} - |\vec{q}+\vec{G}||z|}\,.\nonumber
\end{align}
In the figures of Fig. \ref{fig:fig_2}, \ref{fig:fig_3}, we plot the $z$-component of only the induced field for clarity.

It is instructive to discuss the structure of the equations Eqs. \eqref{eq:phi_tot} and \eqref{eq:etot_snom}. The induced electric field's expression includes the dielectric matrix's inverse.  Thus, it consists of all of the plasmon modes via the determinant of the dielectric matrix formally present in the inverse.  This is similar to the expression for the optical conductivity, Eq. \eqref{eq:optical_conductivity}, however, unlike the optical conductivity that focused only on the $00$ element of the inverse allowing for cancellation of the plasmon zeros in the denominator by the $00$ element of the matrix of cofactors, the expressions Eqs. \eqref{eq:etot_snom},\eqref{eq:phi_tot} involve all elements of the inverse matrix. The plasmon resonances, therefore, are not suppressed. The contribution of the folded branches can, however, be suppressed systematically by controlling the tip distance $z_0$ compared to the parameter controlling the contribution of the off-diagonal elements - the moiré period $L_M$. In the limit of $z_0 \to \infty$, we see (See also Supplemental materials) that the expressions of Eqs. \eqref{eq:etot_snom},\eqref{eq:phi_tot} reduce to the induced potential computed using the macroscopic dielectric function defined previously. This strong dependence on the AFM tip height $z_0$ allows us to turn on/off the contribution of local fields selectively. 

In Fig. \ref{fig:fig_2}, we show the comparison of the expected induced electric field for the two plasmon branches of $\nu=1/2$ antiferromagnet ground state shown in Fig. ~\ref{fig:fig_1}a,b. The notation ``Lower plasmon'' corresponds to the ``conventional'' plasmon mode present in metals, and the ``upper plasmon'' is the folded branch of the plasmon spectrum. Placing the tip such that $z_0  < L_M$ thus enables the inclusion of many terms in the Fourier sum of Eq. \eqref{eq:phi_tot}. The lower and upper plasmon show a distinct structure of electric field profiles, allowing the two branches to be readily distinguished. Specifically, the lower plasmon distribution, aside from a smooth intracell intensity modulation, resembles simply a cylindrical wave profile, see Fig. \ref{fig:fig_2} (upper, middle), seen if local field effects are neglected.
In contrast, the upper plasmon shows a much richer structure of the field modulation with the sign of the electric field mimicking the spin-structure of the AFM order shown in Fig. \ref{fig:fig_1}a. As expected, if local field effects are neglected, Fig. \ref{fig:fig_2} (bottom, middle), the upper plasmon solution does not exist. The last column corresponds to the expected pattern of the electric field signal if only macroscopic dielectric function were used in the calculation or, as discussed above if $z_0 \gg L_M$ such that the contribution of the higher Fourier harmonics is suppressed. The electric field modulation persists and is distinct between the two plasmon branches. Therefore, carrying out an experiment where electric field intensity does not only possess a cylindrical profile, as expected only for the lower plasmon branches, but serves as a direct confirmation of the striped ground-state structure.

\begin{figure*}[t]
\includegraphics[width=\linewidth]{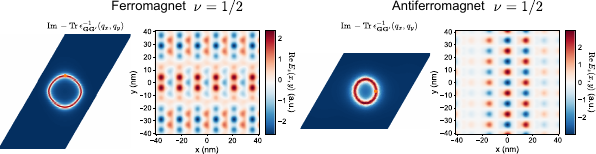}
\caption{Constant energy slices of the trace of inverse dielectric function for filling fraction $\nu=1/2$ and eigenmodes of oscillation obtained through singular value decomposition of $\epsilon_{\mathbf{GG}'}$ for $\mathbf{q}$ indicated by the orange points. In the FM case the modulation on the moir\'e length scale reflects the rectangular unit cell of the Wigner crystal, while in the AFM the pattern is hexagonal in correspondence to the Wigner unit cell in that case.}
\label{fig:fig_4}
\end{figure*}

The plasmon branches' distinct real-space electric field profile persists at larger distances from the AFM tip. In Fig. \ref{fig:fig_3}(a, b), we show the two folded plasmon branches corresponding to the $\nu=1/2$ ferromagnet and antiferromagnet, respectively. Both plasmon modes have an overall cylindrical profile of electric field modulation determined by the BZ momentum $\mathbf{q}$. However, due to the difference in the magnetic order of the two ground states the modulation pattern on the scale of the moiré cell has a different structure. In particular, in the ferromagnetic case within each ring of envelope modulation the moir\'e length scale pattern forms rectangles with vertices at spots of the same sign of $E_z$ (e.g. the red spots), while in the antiferromagnetic case the pattern consists of parallelograms, as also seen in Fig.~\ref{fig:fig_2}. The resulting real-space electric field modulation will therefore be different both between the ground states as well as different plasmon modes (See Supplemental Materials for an extended discussion). This figure also demonstrates that the intricate real-space plasmon structure visible on the scale comparable with the moiré cell is not washed out over larger distances.

The plasmon signal produced by SNOM due to the cylindrical symmetry of the SNOM potential has an envelope of a cylindrical wave. This envelope however is overlaid on a periodic modulation that encodes the eigenvector structure of the electric field that gives rise to the sustained plasmon oscillations, e.g., $\sum_{\vec{G'}} \epsilon_{\vec{G}\vec{G'}}(\vec{q}, \omega) \vec{E}_{\vec{G'}}(\mathbf{q}, \omega) = 0$. As such, the cylindrical wave's long wavelength modulation can obfuscate the plasmon eigenvectors' structure that can encode, for example, the spin order. For this reason, it can be more insightful to consider the SNOM measurement operating in the second mode of operation explained in Sec. IId above. In this experiment, a metallic (typically gold) contact is illuminated and acts as an antenna that can be used to launch plasmons, c.f. Ref. \cite{Woessner2014,10.1038/s41586-018-0136-9}. Carrying out the analysis of the electric potential profile (See Supplemental Materials for derivation), we find that a line-like contact of a length $L$ gives rise to a potential profile which Fourier transform is approximately given by
\begin{align}
\phi_{contact}(q_x,q_y) = 2 \pi \delta(q_x) \times \tilde{\phi}(q_y,L)\,,
\end{align}
where for simplicity, we align the contact with the $x$-axis and $\tilde{\phi}(q,L)$ is a smooth function with a long-tail, i.e. $\tilde{\phi}(q,L) \propto 1/q$ for $q \gg 1/L$ (See Supplemental Materials for a full expression). Correspondingly, see plasmon energy contours in Fig. \ref{fig:fig_4}a for example, we can selectively excite plasmons only with $q_x=0$ and a non-zero $q_y$ component. However, the resulting real-space field structure resembles that of Fig. \ref{fig:fig_2}, without the remnant of the dipole. A plane wave solution, as shown in Fig. \ref{fig:fig_4} (See Supplemental Materials for comparison), corresponds almost exactly to the eigenvectors of the dielectric matrix for the given plasmon mode (shown as an orange point in energy contours of Fig. \ref{fig:fig_4}). This analysis demonstrates the identity between the SNOM or contact-excited electromagnetic wave and the eigenvectors of the dielectric matrix. Moreover, comparing the corresponding profile of the $ z-$ component of the electric field, we find that it mirrors the spin structure of the correlated order, see Fig. ~\ref{fig:fig_1}. As such, we argue that the SNOM, an experimental technique sensitive to the charge density, can, in principle, also operate as a detector of spin order due to the internal structure of the correlated order that couples spin and charge degrees of freedom.

\section{Discussion \& Outlook}

In this work, we showed how the real-space structure of collective charge oscillations of the correlated orders can directly encode information about the internal quantum structure of the ground states. We focused on the problem of generalized Wigner crystals~\cite{tangSimulationHubbardModel2020, reganMottGeneralizedWigner2020, liImagingTwodimensionalGeneralized2021, xuCorrelatedInsulatingStates2020} in moiré transition metal dichalcogenides where the various fractional fillings are believed to host several competing candidate ground states~\cite{panQuantumPhaseDiagram2020, wuHubbardModelPhysics2018,PhysRevB.104.075150,2022arXiv220900664L}. Our analysis builds upon our earlier result\cite{Papaj2023}, where we argued that the presence of a generalized Wigner crystal modifies the plasmon spectrum of the system, giving rise to new collective modes. In that work, however, we have not exhaustively analyzed whether (and how) these collective modes can be excited experimentally. We address this challenge in the current work, arguing that the collective modes, despite remaining gapped at zero momentum, can only be seen using probes that allow for finite momentum transfer between light and the plasmon modes. We focus on scanning near-field optical microscopy to propose how the electric field that comprises various folded plasmon branches can be readily detected. SNOM is fundamentally a charge-sensing-based method, and thus, our results introduce a new regime under which SNOM can also operate as a probe of the spin degree of freedom.

In addition to focusing on possible experimental confirmation of the plasmon folding phenomenon, our work highlights the importance of local field effects in moiré systems. In the current work, as well as in the previous result of Ref. \cite{Papaj2023}, we introduced two concrete analytically tractable toy models to study local field effects and to develop a framework connecting local field contributions to the dielectric function matrix to experimental probes such as conductivity beyond the $\vec{q}\to 0$ limit as frequently discussed in the literature. Our work paves the way for further study and treatment of local field plasmonics in superlattice systems.

\section*{Acknowledgments}
M.P. was supported by the U.S. Department of Energy, Office of Science through the Quantum Science Center (QSC), a National Quantum Information Science Research Center.  M.P. received additional fellowship support from the Emergent Phenomena in Quantum Systems program of the Gordon and Betty Moore Foundation. G. N. was supported by the National Science Foundation (NSF) CAREER award, under award DMR-2145074 G. N. and C.L. were supported by the start-up funds from Florida State University and the National High Magnetic Field Laboratory. The National High Magnetic Field Laboratory is supported by the National Science Foundation through NSF/DMR-1644779 and the state of Florida. 

\section*{References}
\bibliography{references}

\end{document}